\documentclass{llncs}
\usepackage{graphicx}
\usepackage{listings}
\usepackage{amsmath}
\usepackage{beramono}
\usepackage{array}
\usepackage{lmodern}
\usepackage[rightcaption]{sidecap}
\usepackage{wrapfig}
\usepackage[absolute,overlay]{textpos}
\begin{document}
\graphicspath{{./Figures/}}


\title{Camera Obscurer: \\Generative Art for Design Inspiration}
\author{Dilpreet Singh,$^{1}$ Nina Rajcic,$^{1}$ Simon Colton$^{1,2}$ \\ and Jon McCormack$^{1}$ (0000-0001-6328-5064)}
\institute{$^1$SensiLab, Faculty of IT, Monash University, Australia\\
$^2$Game AI Group, EECS, Queen Mary University of London, UK}
\date{}
\maketitle

%
\begin{textblock*}{\textwidth}(4.2cm,24cm) 
  \fbox{%
    \parbox{\textwidth}{\small%
        Preprint of: D.~Singh, N.~Rajcic, S.~Colton and J.~McCormack, `Camera Obscurer: Generative Art for Design Inspiration', in 
        \textit{EvoMUSART 2019: 8th International Conference on Computational Intelligence in Music, Sound, Art and Design}, April 2019, Leipzig, Germany, 2019
          } }
\end{textblock*}

\begin{abstract}
We investigate using generated decorative art as a source of inspiration for design tasks. Using a visual similarity search for image retrieval, the \emph{Camera Obscurer} app enables rapid searching of tens of thousands of generated abstract images of various types. The seed for a visual similarity search is a given image, and the retrieved generated images share some visual similarity with the seed. Implemented in a hand-held device, the app empowers users to use photos of their surroundings to search through the archive of generated images and other image archives. Being abstract in nature, the retrieved images supplement the seed image rather than replace it, providing different visual stimuli including shapes, colours, textures and juxtapositions, in addition to affording their own interpretations. This approach can therefore be used to provide inspiration for a design task, with the abstract images suggesting new ideas that might give direction to a graphic design project. We describe a crowdsourcing experiment with the app to estimate user confidence in retrieved images, and we describe a pilot study where Camera Obscurer provided inspiration for a design task. These experiments have enabled us to describe future improvements, and to begin to understand sources of visual inspiration for design tasks.
\end{abstract}

\section{Introduction and Motivation\label{introduction}}

Producing decorative abstract images automatically can be achieved in numerous ways, e.g., via evolutionary computing \cite{romero:evoartbook} or generative adversarial networks \cite{elgammal:iccc17}. With modern computing power, it's possible to produce tens of thousands of images of fairly high resolution in a short time, and driven by human interaction and/or a suitable fitness function, these images could have high aesthetic appeal and diversity. Such generated images can be incorporated into artistic practice through careful curation and exhibition \cite{reas:formcodebook}, and such images have found other uses, for example as the basis for scene generation \cite{colton:evomusart12}. However, it is fair to say that there are diminishing returns when it comes to the value of large databases of generated abstract decorative images. One of the motivations of the work presented here is to find an interesting usage for large databases of such generated images, namely to provide inspiration for design projects.

Abstract art pieces made by people often offer various opportunities for interpretation and/or self reflection. In the eye of a particular viewer, an abstract image could bear a resemblance to an object, person, animal, texture or place; it could bring back good or bad memories; evoke an emotion; set a mood or reflect certain aesthetic considerations like symmetry, balance, depth and composition. Generated abstract images similarly offer possibilities for interpretation. As people are very good at interpreting meaning in images constructed with none in mind, generated images can often reflect objects, showcase aesthetics, evoke emotions and possibly provide inspiration for tasks.

\begin{figure}[t]
\centering
\includegraphics[width=0.9\linewidth]{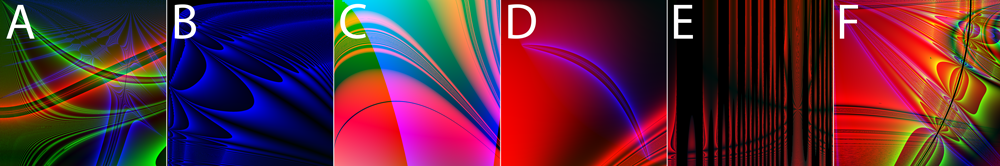}
\vskip -0.15in
\caption{Interpretable decorative images generated by $(x,y) \rightarrow (r,g,b)$ functions.\label{fig:hr3images}}
\vskip -0.1in
\end{figure}

As an example, the images in figure \ref{fig:hr3images} were constructed using a very straightforward approach dating back to Sims \cite{sims:siggraph91}. Here, algorithms which take $(x,y)$ coordinates as input and output triples of $(r,g,b)$ values using trigonometric, arithmetic and exponential functions were automatically generated and run to produce the images. Even though the process for generating these images is simplistic (in generative art terms), the images can still be evocative. For instance, image A may remind a viewer of a complicated motorway junction, while image B may evoke memories of Halloween ghosts. Image C has a look of light reflecting off a shiny surface, while image D looks a little like a feather quill being used on paper. Of images E and F, most people would agree that one could be interpreted as evoking a sinister mood, while the other evokes a fiesta mood.

Some projects have aimed to produce images which invite interpretation in certain ways, e.g., Todd and Latham's Mutator project originally aimed to produce images resembling organic forms \cite{todd:mutator}, and McCormack produced artworks depicting imaginary flower and plants evocative of Australian flora \cite{mccormack:evomusart04}. Machado et. al. have evolved abstract pieces for various interpretations including as faces \cite{correia:iccc13} and figures \cite{machado:evomusart12}. An alternative to generating images for a particular interpretative purpose is to generate large databases of abstract images and use the fact that many will have reasonable visual interpretations (albeit somewhat subjectively). The images in figure 1 were chosen by hand specifically because of places, objects or moods they might evoke in an average viewer. Automating this skill of visually identifying images with valid interpretations could be useful, as it would enable mass-generation of abstract decorative images in an unfocused way, with a user providing focus later on, e.g., looking for images evoking ghosts when the deadline for a Halloween graphic design project is looming. 

An important question when implementing such an approach is how to retrieve abstract images from a large database. There are many potential solutions to this, and we investigate here using visually similar search, where the user supplies a \emph{seed image} for image retrieval, and the app tries to find visually matching images from various datasets. The seed image could come from a variety of sources such as a user's surroundings, a hand-drawn sketch or an image on their computer screen. To test this approach, we have developed a mobile application called \emph{Camera Obscurer} which retrieves images from two large datasets of generated images, given a seed image for a visual similarity search. As described in section \ref{sec:implementation}, the two datasets are: (a) around 9,500 abstract art images similar in nature to those in figure \ref{fig:hr3images} but produced by a different method described below, and (b) 20,000 filtered images, obtained by applying 1,000 generated image filters to 20 digital photos representative of the real world.

Running on a camera-enabled mobile phone or tablet, the system allows the user to take a photograph from their surroundings with their device, instantly receive images from the datasets and upload some to a Pinterest board as they see fit. The first expected usage for Camera Obscurer is simply to find images valuable to a graphic design project, i.e., that might be used in the project. Hence, if a user had a ghost theme in mind for a Halloween design, and wanted an abstract artwork somewhat evocative of ghosts, they could (a) photograph something nearby which looks a little like a ghost (b) draw something ghost-like or (c) bring up an image in their browser of a ghost and photograph that. In principle, if a ghost-like image is available in the database, it should be retrieved. However, while the effectiveness of visual similarity search has been evaluated with respect to retrieving similar digital photographs of real-world images, abstract images don't in general hugely resemble real-world seed images. Hence, to test whether abstract images can be sensibly retrieved, we performed a crowdsource experiment to measure whether retrieved images resemble the seed images more so than a randomly chosen one, as described in section \ref{sec:experiments}.

A welcome side-effect of abstract images having relatively low visual similarities with seed images is that the retrieved images may have numerous visual interpretations, e.g., as objects different from those in the seed image, or with different textures, juxtapositions and/or evoking different moods and emotions. Hence a retrieved abstract image could supplement a user's seed image and provide inspiration. The second expected usage for Camera Obscurer is to provide higher-level inspiration for design projects, over and above the retrieval of potentially useful images. Imagine, for instance, someone undertaking a design project and wanting to employ a texture, but unsure of which to use. Here, they could use the app to photograph textures around them to use as seeds, and inspect any retrieved images resembling textures and hopefully make a more informed choice. This approach may be useful for amateurs undertaking graphic design projects with little design background and/or starting with little or no inspiration.

To investigate the potential of this approach for design inspiration, we undertook a pilot study where 8 participants used Camera Obscurer to help with a design project. To compare and contrast the value of abstract generated pieces, the app also retrieved colour palettes, images from a photo archive and images of human-made art. As described in section \ref{sec:case_study}, the purpose of the pilot study was to raise suitable questions about using abstract retrieved images for design inspiration, rather than to test particular hypotheses. We conclude with these questions in section \ref{sec:conclusions}, and suggest improvements for design tasks by: enabling generative artworks and filters to be changed as part of a design workflow; and enabling the software to provide textual interpretations of abstract pieces.

\section{The Camera Obscurer System\label{sec:implementation}}

\subsection{Image Curation}

To recap, our intention is to test whether abstract generated imagery retrieved in response to a seed image from a user's surroundings could provide visual inspiration. This needs a large dataset of decorative art images which (a) is varied enough to include images having some visual similarity with a large range of seeds and (b) contains images which can be interpreted in various ways. To this end, we used an archive of generated decorative images previously described in \cite{colton:evomusart11} and a dataset of generated image filters, as described in \cite{colton:evomusart09}.

\begin{figure}[t]
    \centering
    \includegraphics[width=0.9\linewidth]{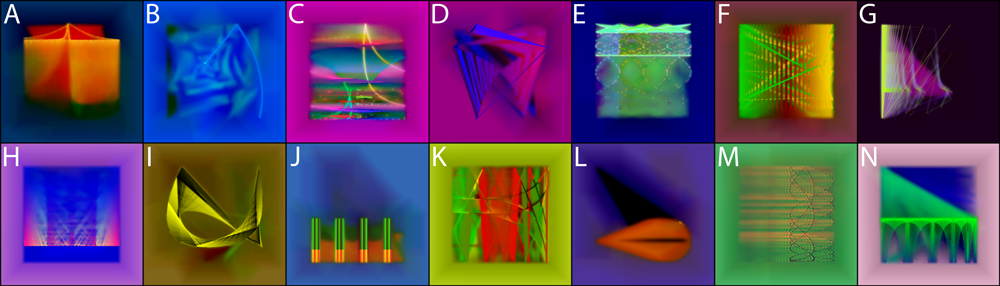}
    \vskip 0.1in
    \includegraphics[width=0.9\linewidth]{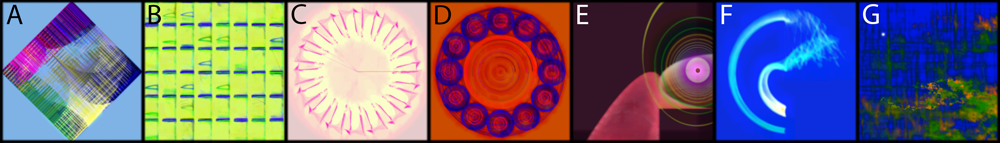}

    \vskip -0.15in
    \caption{(i) Abstract generated images with interpretable depth (B, H, J, L), patterns (E, F, K), motion (C, G, M), lighting effects (C, H, N) , textures (B, E, F, M), objects (D, I, L) and locations (A, C, J, N). (ii) Images produced with (A) diamond (B) grid (C) kaleidoscope (D) necklace (E) oval (F) polar and (G) tron transformations.\label{fig:elvira}}
    \vskip -0.1in
\end{figure}

The decorative images were produced by evolving a series of 10 functions which take as input a particle number, $p$, between 1 and 1,000, a timestep number, $t$, between 1 and 100, and previous values of any of the functions. Each function outputs numerical values which are interpreted as (a) initialisation functions $i_1, \ldots, i_5$ calculating $(x,y)$ coordinates and $(r,g,b)$ colours of 1,000 particles and (b) updating functions $u_1, \ldots, u_5$ calculating the coordinates and colour of a particle at timestep $t$. As each function can use previously calculated values, the calculations performed become highly iterative and complex. To use the functions, an image of a given size is first generated as a monotone background, with the colour chosen randomly. Then 1,000 particles are initialised via $i_1$ to $i_5$ and then moved and changed (in colour) with $u_1$ to $u_5$ over 100 timesteps. At each timestep, a line is drawn between each particle's previous position and its new one, with the line rendered in the new colour of the particle. After each timestep, a blur is also applied to the entire image. 

We have found the images generated in this fashion to contain highly varied, aesthetically pleasing (albeit subjectively) decorative examples, and we have performed a number of experiments to search the space, using fitness functions \cite{colton:evomusart12} and human guidance \cite{colton:evomusart11}. In particular, as per figure \ref{fig:elvira}(i), the generated images can evoke interpretations of depth, patterns, motion, lighting effects, textures, objects and locations, and allow certain moods and emotions to be projected onto the image. Through a number of experiments, we produced around 1,000 high quality images using this method, having hand curated them from around 10,000. To further increase the variety and number of the images produced, we experimented with transformations to the lines before they were rendered on the image. We experimented with 6 transformations (diamond, grid, kaleidoscope, necklace, polar and tron), which moved the start and end points of the lines before rendering them. For instance, with the polar transformation, the Cartesian $(x,y)$ coordinates of the line start and end points were transformed to polar coordinates $(\theta, r)$. This method produces images of a more circular nature. Some transforms had parameters which were also experimented with, e.g., the number of beads in the necklace transformation. We also trialled a transform which replaced the line with an oval with focal points at the line ends. Example images produced using the different transformation are given in figure \ref{fig:elvira}(ii). 

As each transformation produces quite visually different images to each other, we increased the volume of images in the archive by taking the 1,000 curated images and applying the transformation before re-rendering them. However, some transforms spread out the lines on the canvas, producing largely empty, unusable, images. Other transformations took quite busy original images and made them too noisy. In particular, the oval transformation produced a high proportion of visually disorienting and messy images, and we had to remove from the dataset around 60\% of these images, as they did not provide enough coherence for a reasonable interpretation along any of the lines discussed above, so would not be useful for the Camera Obscurer app. Of the other transformations, around 1 in 20 curated images had to be discarded after the transformation. The necklace and kaleidoscope transforms were most successful aesthetically, as their structure provides balance and symmetry not guaranteed by the other transformations. 

\begin{figure}[t]
    \centering
    \includegraphics[width=0.9\linewidth]{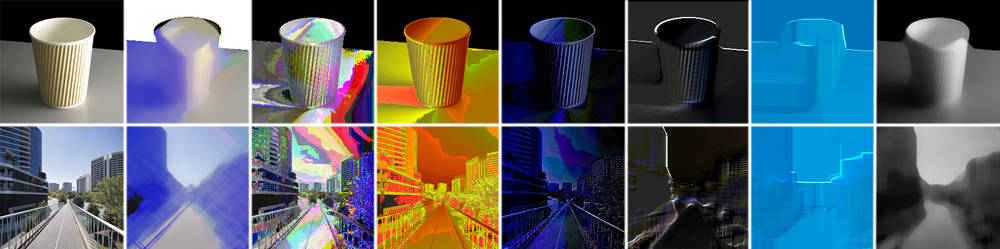}
    \vskip -0.15in
    \caption{Image filter examples: original image on the left, 7 filtered versions on the right.\label{fig:filterfeast}}
    \vskip -0.1in
\end{figure}

We supplemented the 7,000 images which were not removed with 2,500 more, produced in a different run, with no transformation (examples in figure \ref{fig:elvira}(i)). However, in preliminary experiments, we found the border around these images was a substantial visual feature, which meant they were rarely retrieved. Hence, we cropped the border before the visual analysis used for indexing them, but kept the full image available for retrieval. In total, we supplied 9,519 abstract art images to Camera Obscurer. In addition, we supplied 20,000 images obtained by passing 20 images representative of diverse aspects of life through 1,000 image filters. As described in \cite{colton:evomusart09}, the filters were themselves generated via a directed search over a space of trees which pass source images through single transformations like blurring, and also composite two images together, e.g., through a bitwise operator. Many filters produce artistic effects resembling watercolours, embossing, etc., and others produce highly abstracted images. A selection of seven filters applied to two source images is given in figure \ref{fig:filterfeast}. The filters introduce abstraction, new colours, textures and artefacts. For the experiments described below, we also supplied three more datasets, namely: (a) 10,000 images from the \emph{WikiArt} archive (wikiart.org) (b) 10,000 images from the photographic \emph{Archive of the Planet} (http://collections.albert-kahn.hauts-de-seine.fr) and (c) 5,000 thumbnails depicting five colours taken from palettes given in \cite{krause:colour_palettes}.

\subsection{Visual Similarity Image Retrieval}

Content-based image retrieval is a well studied problem in computer vision, with retrieval problems generally divided into two groups: category-level retrieval \cite{sharma2015scalable} and instance-level retrieval \cite{zheng2018sift}. Given a query image of the Sydney Harbour bridge, for instance, category-level retrieval aims to find \textit{any} bridge in a given dataset of images, whilst instance-level retrieval must find \textit{the} Sydney Harbour bridge to be considered a match. Instance retrieval is therefore the much harder problem, and because it has many more applied use cases, it's also the more studied problem.

Traditionally, instance retrieval has been performed using the prominent Bag-of-Words (BoW) model combined with a set of local descriptors, such as scale-invariant feature transforms (SIFT) \cite{lowe2004distinctive}. Other popular approaches use local descriptor and vector aggregation methods such as the Fisher Vector \cite{perronnin2007fisher} and the VLAD representation \cite{jegou2012aggregating}, which have been among the state of the art techniques in this domain. In 2012, Krizhevsky et al. \cite{krizhevsky2012imagenet} demonstrated significant improvements in classification accuracy on the ImageNet \cite{deng2009imagenet} challenge using a convolutional neural network (CNN), exceeding prior results in the field by a large margin. Since then, many deep convolutional neural network (DCNN) architectures have been developed such as VGG, ResNet, and Inception. All of these have pushed forward the state of the art in classification accuracy on the ImageNet challenge. Networks trained on the ImageNet classification tasks have been shown to work very well as off the shelf image extractors \cite{sharif2014cnn}, and show even better results when fine-tuned to datasets for the task at hand \cite{azizpour2015generic}. DCNNs are also commonly used as \textit{backbone} networks in more complex multi-stage architectures (functioning as the base-level image feature extractors) which led to state-of-the-art performance in tasks like object detection \cite{lin2018focal} and object segmentation \cite{he2017mask}.

Experiments in applying CNNs in the field of image retrieval have shown promising results \cite{babenko2014neural}, in some cases surpassing the performance of traditional methods like SIFT \cite{zheng2018sift} and VLAD \cite{azizpour2015generic}. Commercial organisations such as Bing and Pinterest have completely moved away from hand engineered features in their image retrieval pipelines, as described in \cite{hu2018bing} and \cite{jing2015pinterest} respectively. Instead, they have adapted one or more DCNN architectures as a way of creating compressed image representations. The evaluation of these systems primarily consists of applying them to datasets with real-life photographs containing easily recognisable objects and/or scenes, with concrete shapes and form, and with inherently natural colour illumination. In contrast, the kinds of computer generated abstract art datasets that we are concerned with, exhibit very few of these naturalistic properties, and differ significantly in terms of their characteristics. 

Research in the field of image retrieval, both traditional and CNN based, has not often included evaluating retrieval over abstract images. More diverse datasets have been used in tasks around identifying art genres \cite{zujovic2009classifying}, style \cite{bar2014classification}, and predicting authors of an artistic work \cite{hicsonmez2017draw}, but these are in the domain of classification rather than retrieval. Seguin et al. \cite{seguin2016visual} evaluated image retrieval of artworks in the historical period from 1400 to 1800, and showed that BoW with SIFT descriptors didn't transfer well to paintings. They showed that a pre-trained VGG performed well, but they achieved their best results by fine-tuning VGG on their dataset. Performance of VGG on capturing style and content has been well demonstrated by Gatys et al. \cite{gatys2016neural}, demonstrating that a pre-trained VGG on a non-domain specific dataset can capture features which identify style and content of artworks.

When implementing retrieval systems for more practical purposes, accuracy is not the only dimension worth measuring. In addition, aspects such as search efficiency, open-source availability, and device support are major considerations. Apple and Google have optimised both their mobile operating systems to natively support running of complex neural network models, offering Core ML and ML Kit respectively. This level of native support combined with the larger open source community around deep learning played a major role in our choice of only evaluating DCNNs for the feature extraction abilities in the Camera Obscurer app. We selected architectures from the three families: VGG-16, ResNet-50, and Inception-v4. All of these networks were downloaded with pre-trained ImageNet weights and used as off-the-shelf feature extractors. For each network, the last non fully-connected layer of the network was used as the feature representation, as it tends to provide a good starting point \cite{razavian2016visual}. We discard the fully-connected layer(s) as they are ImageNet specific and not relevant for feature representation. This meant that for a single image, VGG-16 and ResNet-50 had 2048 features, and Inception-v4 contained 1536 extracted features.

\begin{figure}[t]
\centering
    \fbox{
    \includegraphics[width=0.9\linewidth]{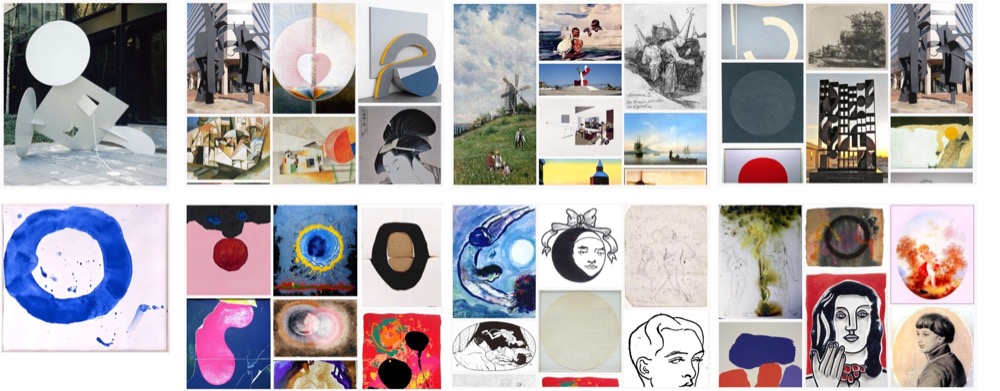}
    }
    \vskip -0.1in
    \caption{Examples from the 3 networks for the given seed images (leftmost). In general, ResNet-50 (2nd column) performs much better than both VGG-16 (3rd column) and Inception-v4 (4th column) at visual similarity image retrieval for abstract image seeds.
    \label{fig:network_comparison}}
    \vskip -0.1in
\end{figure}

We carried out an empirical evaluation to compare how each network performed on visually similar image retrieval. A sample of the WikiArt corpus was chosen for the evaluation, as it contains a range of different image styles. The dataset contained 10,000 artworks from differing style categories to provide a mix: Realism (2000), Pointillism (500), Contemporary Realism (300), New Realism (200), Impressionism (2000), and Abstract Expressionism (2000). All three networks performed well at retrieval tasks that related to realistic paintings, as expected, but ResNet-50 performed best with respect to more abstract images, as portrayed in figure \ref{fig:network_comparison}, which shows favourable examples from all three networks. In a subjective visual inspection, we found ResNet-50 to be the network that reliably returned the most coherent results, i.e., images that were clearly more visually similar and sensible than those returned by the other networks. In terms of search efficiency, ResNet-50 is also the quickest for inference, requiring in the order of 4 times fewer operations than VGG-16. We were able to run feature extraction using the network at a rate of 15 images per second on an iPhone X using Apple's Core ML toolchain, which is sufficient for our use-case.

\begin{figure}[t]
\centering
\fbox{
    \includegraphics[width=0.9\linewidth]{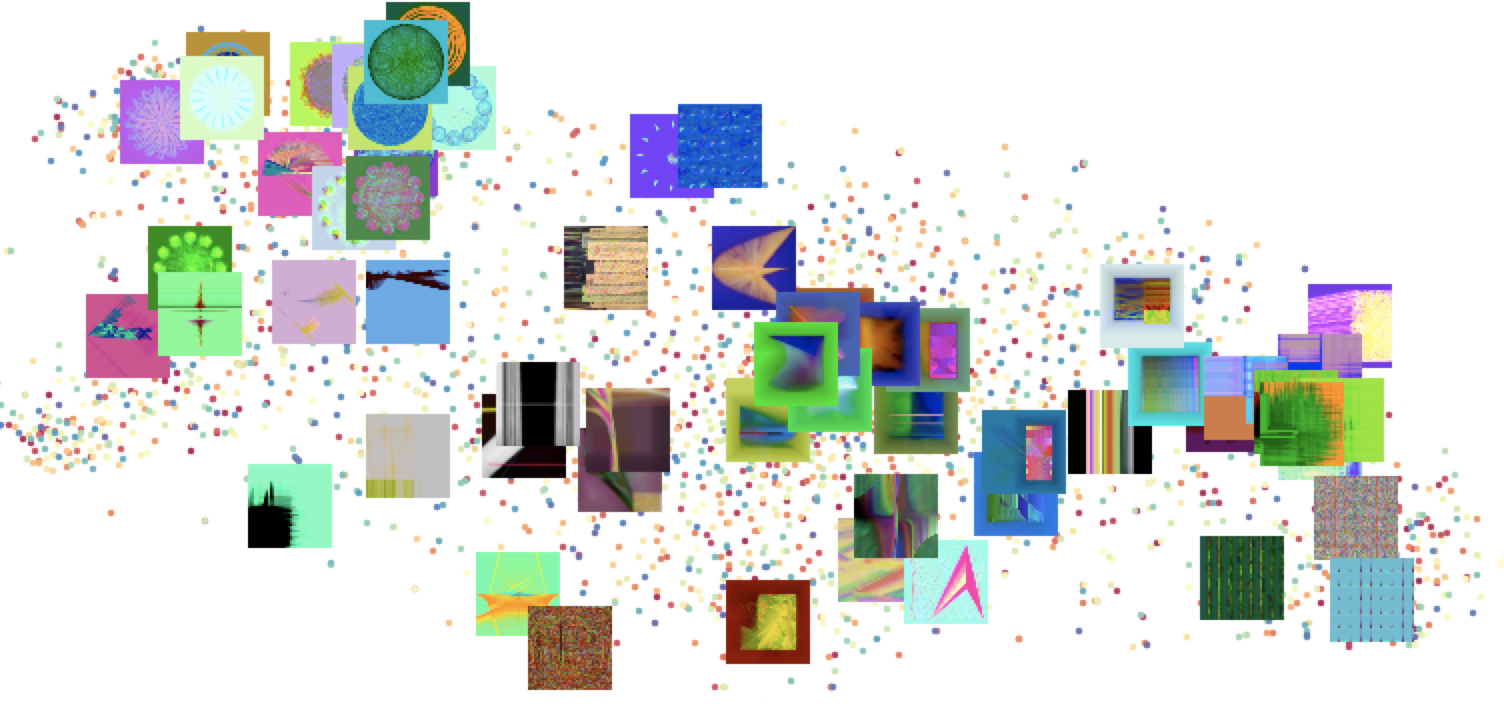}}
    \vskip -0.1in
    \caption{T-SNE plot showing clusters of visually similar images. The plot shows that a ResNet-50 functions well as a generic feature extractor for highly abstract images.
    \label{fig:tsne}}
    \vskip -0.1in
\end{figure}

As the WikiArt dataset contained no computer generated abstract images, we further evaluated the performance of ResNet-50 on the dataset containing 9,519 completely abstract images described above. As it would be difficult to construct an objective test-set for evaluating retrieval on these images, we instead evaluate the network extracted features. Figure \ref{fig:tsne} shows a 2-dimensional projection of a subset of the features (3000 samples) constructed using T-SNE (a popular dimensionality reduction technique). For visual clarity, only 2\% of the sample images are shown on the plot, which displays clear image clusters with similar elements: images with circular aspects towards the top left, with matrix-like patterns towards the right. T-SNE doesn't preserve any kind of distance measure, thus no assessments can be made regarding which two groups are the furthest in embedding space. The figure clearly indicates that the ResNet-50 neural network can function as a generic feature extractor even for the most abstract images. Whilst DCNNs may not be the best at representing abstract art for deeper comparisons \cite{wang2013som}, their attributes such as the off-the-shelf nature, fast inference, and automatic embedding made them a valid choice for Camera Obscurer.

With this embedding representation of an image, the next step in image retrieval is to find and retrieve similar search results. Computing exact nearest neighbours is feasible in evaluation datasets, but as the number of images increases to the high hundreds of thousands, it becomes infeasible to guarantee exact search, especially when search efficiency is a priority. Spatial indexes like k-d trees try to address the problem of exhaustive search by using binary trees to divide up the embedding space, but they only work for feature vectors containing 10s of dimensions, not hundreds or thousands. To address this curse of dimensionality, we turn to \textit{Approximate} Nearest Neighbour (ANN) search algorithms. Methods such as Locality Sensitive Hashing (LSH), Randomised Partition Trees (RPT) don't guarantee exact nearest neighbours, but do promise extremely fast results in very close neighbourhoods of the query.

\begin{figure}[t]
\centering
    \includegraphics[width=0.96\linewidth]{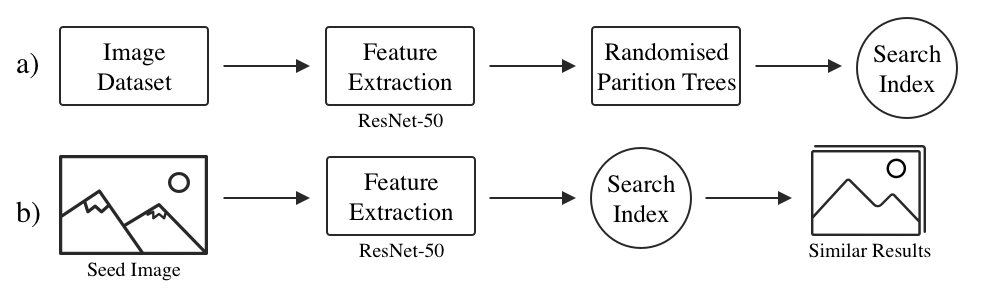}
    \vskip -0.2in
    \caption{Architecture of the image retrieval system.
    \label{fig:architecture}}
    \vskip -0.2in
\end{figure}

We used an implementation of RPT combined with random projections that has shown to have competitive performance for both retrieval accuracy and memory usage \cite{aumuller2017ann}, and is used by the popular music streaming service Spotify for music recommendations (see https://github.com/spotify/annoy). Cosine distance (proportional to euclidean distance of normalised vectors) was applied as the metric when constructing the search index as it is insensitive to vector magnitude. The initial search index is computed off-device as it is computationally expensive, but subsequent search queries (on an index containing 50,000 images) can be performed on the device in less than 50ms. The query time scales logarithmically as the underlying spatial index is constructed using binary search trees, so the search performance isn't impacted too heavily as the dataset grows. Further improvements to the retrieval accuracy were made by using a forest of RPTs, which incurs minimal cost in search efficiency as queries can be executed in parallel, but increases the likelihood of an exact match, as described in \cite{dasgupta2013randomized}.

The architecture of our image retrieval system is shown in Figure \ref{fig:architecture}. The top half shows the construction of the search index from an image dataset using a ResNet-50 feature extractor combined with RPT based ANN search index. The bottom half portrays how subsequent search queries are computed. Our implementation is sufficiency modular, so that both the feature extractor and the search method can be updated when needed. However, with our current choices, we achieve practically real-time performance on a mobile device.

\subsection{User Interface}

\begin{figure}[t]
\centering
\includegraphics[width=0.9\linewidth]{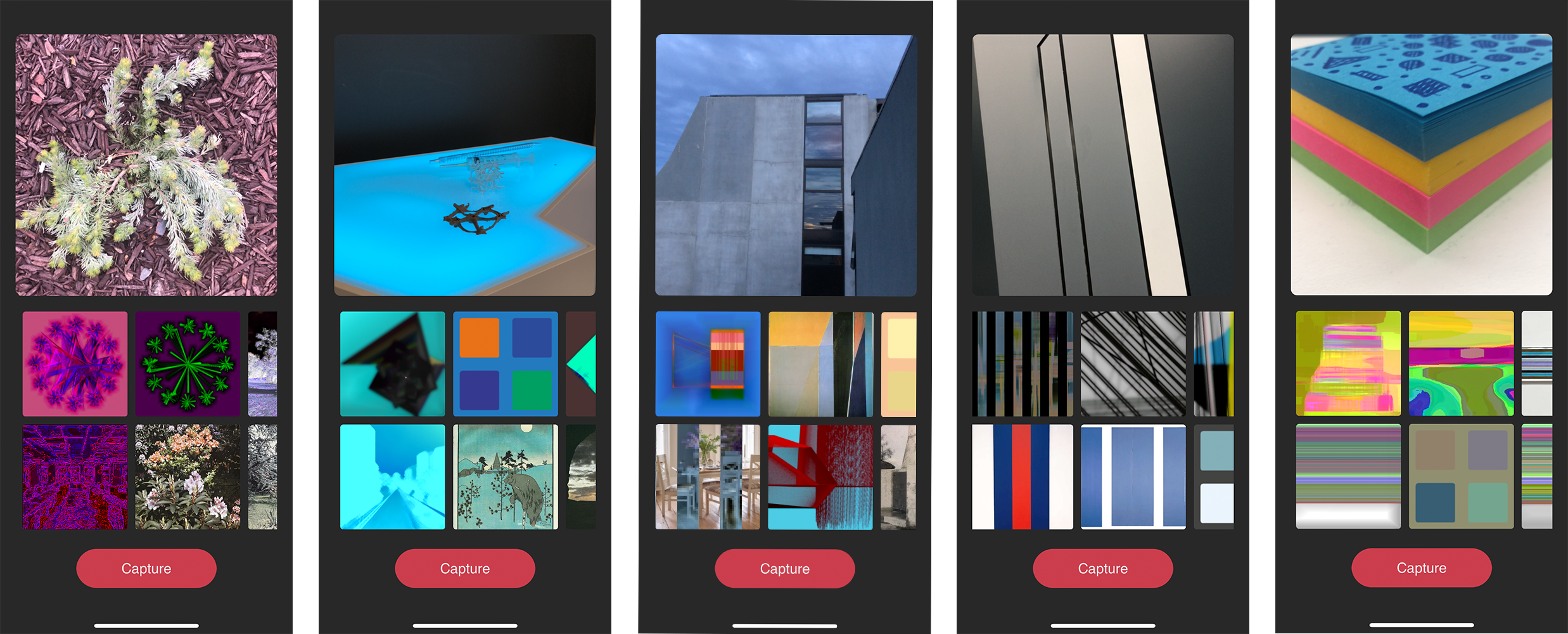}
\vskip -0.1in
\caption{Screenshots from the Camera Obscurer app.\label{fig:screenshots}}
\vskip -0.1in
\end{figure}

The Camera Obscurer app is still an early prototype, developed initially for handheld iOS devices. It has a minimal interface consisting of a large video box at the top of the screen -- which streams a feed live from the camera on the back of the device -- and 10 image boxes below this which show images retrieved from the datasets. It also has a capture button at the bottom of the screen which initiates the process of taking the current still from the video feed as the seed for the visual similarity search. A user can also tap the video box to start the retrieval process, or tap any of the 10 retrieved images, in which case the app uses that as the seed image for a new search. The user can further double tap on any box (image or video) to pin the shown image to a Pinterest board. They can also shake the app in order to replace the currently retrieved images with the previous set. Screenshots from the Camera Obscurer app are given in figure \ref{fig:screenshots}, highlighting the kinds of image retrieval performed for given seed images taken from local surroundings, both indoors and outdoors. 

\section{Evaluation}

In the case of a standard visual similarity search within a comprehensive image database, the method of evaluating the search algorithm would normally involve pixel-to-pixel matching between the seed image and the retrieved image \cite{Keysers04pixel-to-pixelmatching}. In our case, however, the image database is intentionally restricted as the retrieved images are intended to serve as a source of visual inspiration, rather than provide an exact match. For this reason, pixel-to-pixel matching is not a suitable measure of success. Moreover, in our evaluation of Camera Obscurer, we wish to investigate two hypotheses: not only that the retrieved images bear some perceived visual similarity to the seed image, but in addition, that the retrieved images can serve as a source of visual inspiration in the design process. The experiment and pilot study below address these aspects respectively.

\subsection{A Visual Similarity Experiment\label{sec:experiments}}

To measure the perceived visual similarity between a seed image and a retrieved image, we devised a crowdsource experiment utilising Amazon's Mechanical Turk service (mturk.com). 100 participants were presented with a seed image (taken from the MIT-Amobe FiveK Dataset \cite{fivek}) accompanied by two retrieved images, namely: one image retrieved through our visual similarity search, and the other retrieved randomly. Participants were then asked to select the retrieved image with strongest visual similarity to the seed, basing their decision on colour, mood, texture, shape and content. The seed image dataset was chosen as it provides a suitable representation of the style of images we expect to be taken by future users of Camera Obscurer. Each participant was asked to select the most visually similar image for 100 seed images across three datasets. These comprised 32 seed images from each of: (a) the abstract generated images (figure \ref{fig:elvira}) (b) the filtered images (figure \ref{fig:filterfeast}) and (c) the WikiArt dataset (wikiart.org). The remaining 4 seed images were accompanied with one exact match to the seed image and one randomly selected image, to serve as control questions.

\begin{table}[t]
\centering 
\begin{tabular}{|c|c|c|c|c|c|c|c|} 
\hline
Dataset & Avg Acc. & Avg Dist. & Pearson Coefficient & Q1(\%) & Q2(\%) & Q3(\%)& Q4(\%)\\
\hline 
WikiArt & 90.91\% & 0.423 & 0.23 & 25.0 & 25.0 & 20.8 & 62.5\\
Filtered & 85.56\% & 0.477 & 0.17 & 37.5 & 41.7 & 33.3 & 20.8\\
Abstract & 84.89\% & 0.568 & 0.17 & 37.5 & 33.3 & 45.9 & 16.7\\
\hline
All & 87.12\% & 0.490 & 0.22 & 100 & 100 & 100 & 100\\
\hline 
\end{tabular}
\vskip 0.1in
\caption{Summary results across three image datasets, displaying average accuracy, average cosine distance between seed and retrieved image, Pearson correlation coefficient between average cosine distance and count of incorrect selections, and the percentage population the datasets in each accuracy quartile.\label{table:mturk}}
\vskip -0.2in
\end{table}

As shown in table \ref{table:mturk}, the intelligently retrieved images had a greater perceived visual similarity than the randomly retrieved images, as participants correctly selected the intelligently retrieved image with an accuracy of 87.12\%, averaged over the 96 (non-control) seed images. Using a binomial test, we were able to reject the null hypothesis that participants were as likely to select the random image (as being the closest visually to the seed) as they were to select the intelligently retrieved one ($p < 0.001$). The dataset with the highest accuracy was WikiArt, with the intelligently retrieved images being selected 90.91\% of the time. The abstract generated images and filtered image datasets achieved 84.88\% and 85.55\% accuracy respectively. The average time of appraisal among participants was 8.71 seconds per image. To investigate labelling accuracy per dataset, we present the distribution of images in each dataset across accuracy quartiles in table \ref{table:mturk}. We see that WikiArt images account for 62.5\% of the images with highest perceived visual similarity (Q4), whereas the filtered and abstract images combined account for 75\% of the images with the lowest perceived visual similarity (Q1 and Q2). 

The cosine distance between seed and retrieved image appears to be mildly positively correlated ($r = 0.22$) with the number of incorrect selections made, as per the Pearson product moment coefficients in table \ref{table:mturk}. The correlation is higher for the WikiArt images, and lower for the filtered and abstract images. This was to be expected given the higher abstraction levels exhibited in the abstract art and filtered images. For illustrative purposes, in table \ref{tab:example_images}, three example questions from the crowdsourcing test are given, along with the labeling accuracy and the cosine distance. As expected, as the distance of the retrieved image from the seed increases, the crowdsourced labelling accuracy reduces.

{\centering
\begin{table}[t]
\centering
\begin{tabular}{|c|c|c|c|c|c|}
\hline
Dataset & Seed & Retrieved & Random & Accuracy & Distance\\
\hline
\raisebox{0.45in}{WikiArt} &
\begin{minipage}[t]{0.15\linewidth}
\includegraphics[width=\columnwidth]{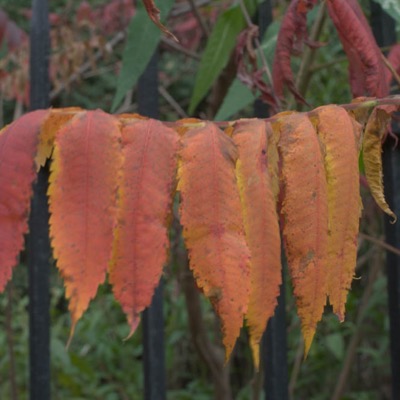}\label{fig:taba}
\end{minipage}
&
\begin{minipage}[t]{0.15\linewidth}
\includegraphics[width=\columnwidth]{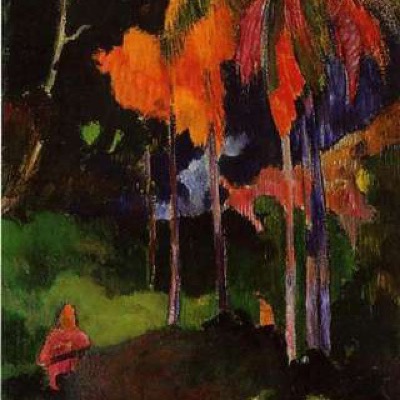}\label{fig:tabb}
\end{minipage}
&
\begin{minipage}[t]{0.15\linewidth}
\includegraphics[width=\columnwidth]{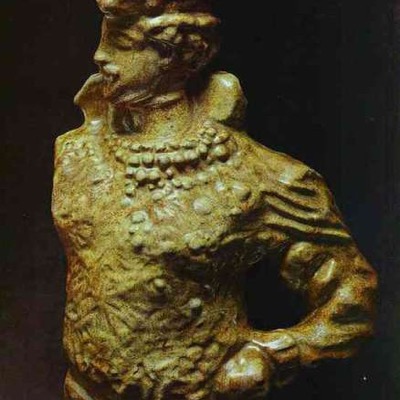}\label{fig:tabb}
\end{minipage}
&
\raisebox{0.45in}{100.0\%} &
\raisebox{0.45in}{0.452} \\
\hline
\newline
\raisebox{0.45in}{Filtered} &
\begin{minipage}[t]{0.15\linewidth}
\includegraphics[width=\columnwidth]{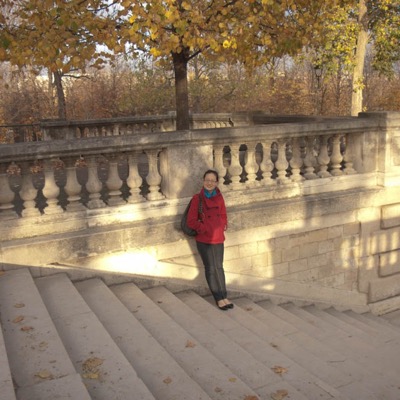}\label{fig:tabc}
\end{minipage}&
\begin{minipage}[t]{0.15\linewidth}
\includegraphics[width=\columnwidth]{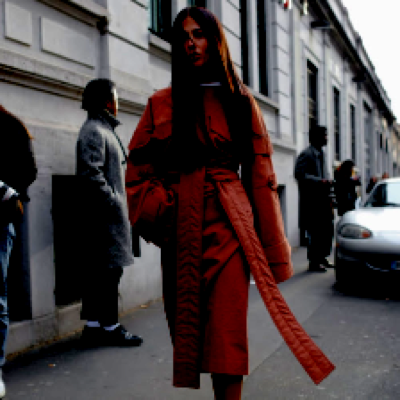}\label{fig:taba2}
\end{minipage}&
\begin{minipage}[t]{0.15\linewidth}
\includegraphics[width=\columnwidth]{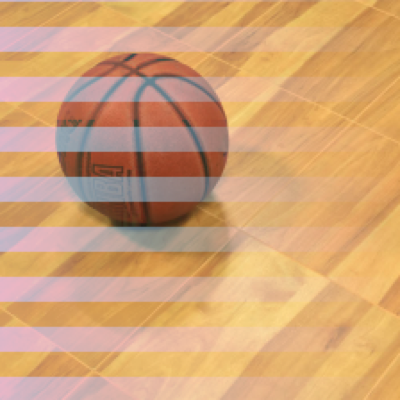}\label{fig:taba2}
\end{minipage}&
\raisebox{0.45in}{92.86\%} &
\raisebox{0.45in}{0.470} \\
\hline
\newline
\raisebox{0.45in}{Abstract} &
\begin{minipage}[t]{0.15\linewidth}
\includegraphics[width=\columnwidth]{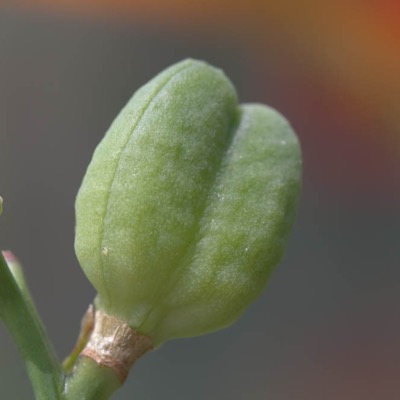}\label{fig:tabc}
\end{minipage}&
\begin{minipage}[t]{0.15\linewidth}
\includegraphics[width=\columnwidth]{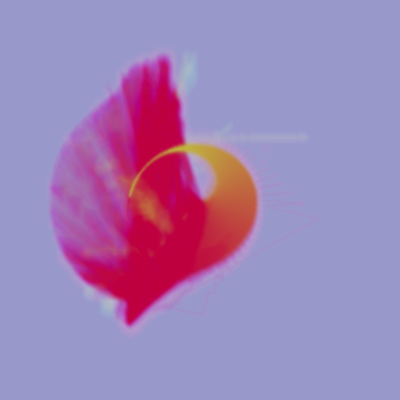}\label{fig:taba2}
\end{minipage}&
\begin{minipage}[t]{0.15\linewidth}
\includegraphics[width=\columnwidth]{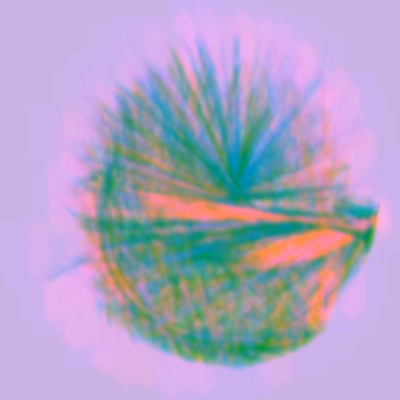}\label{fig:taba2}
\end{minipage}&
\raisebox{0.45in}{32.65\%} &
\raisebox{0.45in}{0.534} \\
\hline
\end{tabular}
\vskip 0.1in
\caption{Example images from each dataset, with accuracy and distance values.\label{tab:example_images}}
\vskip -0.2in
\end{table}
}

\subsection{A Design Inspiration Pilot Study\label{sec:case_study}}

The crowdsourcing results above indicate that abstract images retrieved by the Camera Obscurer app share enough visual similarity with the seed images to be easily distinguishable from images retrieved at random. We felt that this was a sufficient basis on which to carry out a pilot study into the affordances of the app with respect to visual inspiration. To this end, we asked 8 participants to undertake a 3 stage session, for us to gather initial results about app usage. 

In stage 1, participants were given a mock design task which involves designing an album cover for an upcoming ambient electronic musician, whose music describes ``scenes of sprawling concrete structures juxtaposed with vivid internal landscapes of the creatures inhabiting them". In stage 2, participants were given up to 30 minutes to employ Camera Obscurer to take photos from their surroundings with which to retrieve images from the five datasets. They were asked to pin any image (from the camera and/or the retrieved images) that they felt provided inspiration for the design task to a Pinterest board, but were given the guidance of pinning none, if they saw fit. Participants were told that they could tap the retrieved images to use them as a seed, and could shake the device to re-load previously retrieved images. In stage 3, participants were interviewed about their proposed design, and asked specifically how images pinned to their board inspired their final design idea, if at all.

\begin{table}[t]
    \centering
    \begin{small}
    \begin{tabular}{|c|c|cccc|cccccc|}
    \hline
    Part. & Duration & Ext.Seeds & Int.Seeds & Pins & Yield & Abs. & Arch. & Cam. &  Fil. & Pal. & WikiArt \\
    \hline
    1 & 5m41s & 19 & 5 & 20 & 0.83 & 2 & 3 & 10 & 4 & 0 & 1 \\
    2 & 17m56s & 42 & 13 & 31 & 0.56 & 5 & 3 & 13 & 3 & 3 & 4 \\
    3 & 12m32s & 39 & 7 & 34 & 0.74 & 4 & 2 & 17 & 5 & 0 & 6 \\
    4 & 5m36s & 2 & 15 & 16 & 0.94 & 3 & 2 & 1 & 7 & 0 & 3 \\
    5 & 17m05s & 30 & 5 & 21 & 0.60 & 5 & 1 & 5 & 3 & 1 & 6 \\
    6 & 20m37s & 29 & 14 & 59 & 1.37 & 17 & 8 & 7 & 9 & 2 & 16 \\
    7 & 10m13s & 15 & 52 & 21 & 0.31 & 3 & 3 & 0 & 9 & 0 & 6 \\
    8 & 8m53s & 11 & 2 & 14 & 1.08 & 1 & 3 & 3 & 3 & 1 & 3 \\ \hline
    Avg & 12m19s & 23.38 & 14.13 & 27.00 & 0.72 & 5.00 & 3.13 & 7.00 & 5.38 & 0.88 & 5.63 \\
    \hline
    \end{tabular}
    \end{small}
    \vskip 0.1in
    \caption{Summary results for the design inspiration sessions.\label{table:pilot_study}}
    \vskip -0.4in
\end{table}

Summary results from the 8 sessions with (Part)icipants are given in table \ref{table:pilot_study}. We see that the average duration of the sessions was around 12 minutes, and that, on average in each session, 23 images from (Ext)ernal surroundings were used as seeds, while 14 retrieved (Int)ernal images were used as the seed. Participants pinned 27 images on average per session, and we calculated a yield as the number of images pinned per retrieval action, with table \ref{table:pilot_study} showing that around 3 images were pinned for every 4 retrievals. The breakdown of pinned images per archive shows that images taken directly from the (Cam)era were the most likely to be pinned (on average 7 per session), with images from the (Abs)stract art, (Fil)tered images and WikiArt equally likely to be pinned next, roughly (around 5 per session). Images from the (Arch)ive of the Planet were more rarely pinned (3 per session), and the colour (Pal)ette images were very unpopular, with less than one on average being pinned per session. Participants 4 and 7 were outliers in the sense of using the internal (retrieved) images more often than the external images as seeds.

In the stage 3 interviews, we endeavoured to determine if the retrieved images had in any way provided inspiration for the design project that participants believed they would not have otherwise gained. One of the aims of the pilot study is to begin to determine how we might measure visual inspiration more concretely, hence we were broad in our assessment of inspiration. We also asked whether any retrievals were unexpected, and if this effected their ability to generate ideas.

A number of trends emerged from the questioning. Firstly, most participants expressed some frustration with the app design, in particular that they couldn't view the images that they had already pinned. We found that if participants started out with a specific design idea in mind, the app could be disruptive to their creative process, particularly for those with a background in design. However, participants that had no previous design experience found that the app actually allowed them to form their design intention. Many participants describe a `moment' at which a retrieved image sparked an idea for their design. A number of participants enjoyed the abstract nature of the retrieved images, rather than when the app returned a ``literal match", e.g., a participant stated that ``when taking a photo of grass, and was returned a photo of grass, I didn't really find it interesting". In addition to design inspiration, one participant expressed that the app helps to ``capture unconscious feelings" associated to their surroundings.

\section{Conclusions and Future Work\label{sec:conclusions}}

We have implemented and tested an approach which has the potential to provide instant visual inspiration when desired in graphic design tasks. The approach involves image retrieval based on visual similarity, and supplies users with abstract art pieces, filtered images, colour palettes, historical photographs and images of human artwork in response to an image taken from their surroundings. We performed a crowdsourcing experiment to test whether the image retrieval performs adequately when applied to the retrieval of abstract images, and we found that this was the case. In addition, we undertook a pilot study into the value of the app in a design scenario, to explore whether it had the potential to provide visual inspiration. We found that -- even though it is an early prototype -- people were able to employ it in their process with good effect, as on occasion it gave them ideas they probably wouldn't have thought of otherwise.

From the pilot study, we can put forward a number of hypotheses to be tested with a larger empirical study: (i) users pin more images for later reference directly from the camera than retrieved from any dataset (ii) abstracted retrieved images, whether of human art or generative art, are pinned more often for design inspiration than images from reality (iii) graphic design tasks can be influenced by abstract artistic images retrieved in response to visual stimuli, through visual inspiration afforded by the app. To test these hypotheses, we will need to define a more concrete notion of what visual inspiration is in a design setting. Our current thinking is that this needs to be recorded at the time of inspiration, and we expect to implement voice recording as both a design tool for users (i.e., to record their thoughts while they use the app), and as an experimental tool which will enable us to pinpoint moments of visual inspiration. 

We also plan to improve the app to take onboard some of the criticisms which arose during the pilot study. Our aim is to keep the interface as simple as possible, while allowing users more requested functionality, such as viewing the images that they have already pinned. One interesting suggestion that we plan to explore is to enable users to choose image datasets in advance and/or upload their own archive for a particular task. In this way, when using Camera Obscurer, say, for a fashion design task, they could upload datasets of garments, fabrics, etc. As indexing the images can be done very quickly, this functionality is definitely a possibility. Noting also that the retrieved abstract art images can be altered/evolved and filtered images have filters which can be altered and employed too, we plan to enable Camera Obscurer to be part of a design workflow where processes are retrieved for in-app usage, rather than just static images.

The work described here is related somewhat to that in \cite{machado:ijcaiworkshop}, where images were evolved specifically to be multi-stable computationally, i.e., ambiguous to machine vision systems, hence analagous to human interpretation of ambiguous imagery. Our approach is also to supply visually ambiguous images to users, but ones which have some visual similarity with a seed image. The ambiguity is a side-effect of matching abstract artistic images to photo-realistic images, and serves as the basis for visual inspiration. Capitalising on such ambiguity, we plan to use similar retrieval methods as above to find ImageNet \cite{deng2009imagenet} images with abstract art images as the seed. In this way, Camera Obscurer may be able to provide invented texts to accompany the abstract images it retrieves, e.g., for the centre image at the bottom of table \ref{tab:example_images}, it might find a match with images of birds in ImageNet, and invent the tag: ``red bird in flight", perhaps employing some natural language processing technologies to finesse the output. We plan to test whether this leads to further design inspiration, and in general to explore ways in which AI systems can work as inspirational partners alongside designers. We hope to show that generated abstract art can stimulate design thinking, providing much inspiration for artists and designers.

\section*{Acknowledgements}

We would like to thank the participants in the pilot study for their time and energy, members of SensiLab for their very useful feedback on the Camera Obscurer app, and the anonymous reviewers for their helpful 
comments.

\end{document}